\documentclass[manuscript]{aastex}

\usepackage{graphicx}
\usepackage{natbib}

\def \phcmsec{\hbox{photons$\,$cm$^{-2}$s$^{-1}$} }

\def \gray {$\gamma$-ray }

\def \source {\hbox{3C 454.3 }}
\def \cc {3C 454.3 }
\def \ccp {3C 454.3\/}

\def \mtt { }
\def \mttt { }

\def \mta { }

\def \vv {  }

\def \v { }

\def \vvv { }

\slugcomment{Submitted to the \textit{Astrophysical Journal}:
February 22, 2014. Accepted: July 19, 2014.}

\shorttitle{The blob crashes the mirror}

\shortauthors{Vittorini et al. 2013}

\begin{document}

\title{The blob crashes into the mirror:
\\ modelling the exceptional \gray flaring activity of 3C 454.3 in November 2010}
\vspace{2cm}
\author{ V.~Vittorini\altaffilmark{1},
M.~Tavani\altaffilmark{1,2,3}, A.~Cavaliere\altaffilmark{2},
E.~Striani\altaffilmark{2}, S.~Vercellone\altaffilmark{4}}
\altaffiltext{1}{INAF/IASF--Roma, Via del Fosso del Cavaliere 100,
I-00133 Roma, Italy} \altaffiltext{2}{Univ. ``Tor Vergata'', Via
della Ricerca Scientifica 1, I-00133 Roma, Italy}
\altaffiltext{3}{Gran Sasso Science Institute, viale Francesco
Crispi 7, I-67100 L'Aquila, Italy}
 \altaffiltext{4}{INAF/IASF--Palermo,
Via Ugo La Malfa 153, I-90146 Palermo, Italy}
 \altaffiltext{*}{Email: \texttt{vittorini@roma2.infn.it}}

\begin{abstract}

\cc is a  prominent Flat Spectrum Radio Quasar that in recent
years attracted considerable attention because of its variable
high-energy emissions. In this paper we focus on the exceptional
flaring activity of \cc that was detected by AGILE and by
Fermi-LAT in November, 2010. In the light of the time varying data
ranging from the radio, optical, X-ray up to GeV \gray bands, we
discuss a theoretical framework  {\mttt addressing all data } in
their {\mta overall} evolution.
{\vv For two weeks} the source has 
shown a plateau of enhanced GeV emission preceding a {\mta sudden}
major flare lasting about 3 days before decaying. The $\gamma$-ray
flare onset is {\mta abrupt} (about 6 hours), and {\mta is}
characterized by  a prominent "Compton dominance" with the GeV
flux exceeding the pre-flare values by a factor of 4-5. {\mttt
During this episode, the optical and X-ray fluxes increased by a
factor around $ 2$.}

Within  the standard framework of a jet launched with a Lorentz
bulk factor $\Gamma \sim 10$ from a central black hole, we {\vv
explore the yields of} two alternatives. Case 1, with high-energy
emission originating within the broad line region
(BLR); and Case 2, with most of it 
produced outside,  at  {\vv larger distances of a few parsecs}. We
show that Case 1 has considerable problems in explaining the whole
set of multifrequency data. Case 2, {\mta instead},  leads to a
consistent and interesting interpretation based on {\vv the}
enhanced inverse Compton radiation {\vv that is} produced as the
jet  crashes onto a mirror cloud positioned at parsec scales. This
model explains the $\gamma$-ray vs. optical/X-ray behavior of \ccp
, including {\vv the otherwise puzzling phenomena such as the
prominent "orphan" optical flare, and the} enhanced line
emission
with no appreciable $\gamma$-ray 
{\mta counterpart} that preceded the GeV $\gamma$-ray flare. It
also accounts for the delayed onset of the latter on top of
the long plateau.
Our modelling of the exceptional \cc $\gamma$-ray flare shows that,
while emission inside the canonical BLR is problematic,
major {\vv and rapid variations} 
can be produced at  
parsec scales   
with moderate bulk Lorentz factors $\Gamma\approx 15$. Our "crashed
mirror" model {\vv is also applicable}
to other blazar {\mta flares} 
marked by 
{\mta large} Compton dominance of the {\mta emitted radiation}
{\mttt such as  the flare of PKS 1830-211 in}
October 2010.


\end{abstract}

\keywords{gamma rays: observations ---  FSRQ objects: individual
\source}

    \section{Introduction}

Blazars are {\mta powered by } massive black holes (BH) which
launch relativistic jets  with {\mttt considerable} bulk Lorentz
factors $\Gamma \sim 10$ that are aligned with our line of sight.
The variability of their flux radiated in different
energy bands 
{\mta  provides} a promising avenue to understand the physical
processes involved in particle acceleration and in the high energy emissions from 
{\mta such sources.}


{\vv Blazars of the Flat Spectrum Radio Quasar (FSRQ) type are
characterized by beamed continuum emission in the IR-optical-UV
bands of synchrotron radiation nature and also show broad optical
 emission lines (e.g., \citealp{peterson, netzer}). Variations
{\vv of the flux in the optical continuum are often observed to be
correlated and simultaneous with flux variations in higher energy
bands. This indicates} a key role in the \gray emission played by
inverse Compton scattering of surrounding seed photons off the
same {\mta highly} relativistic electrons that radiate the
synchrotron. Adequate seed photons can be supplied {\mta not only}
by the beamed synchrotron emission itself {\v (the so called Synchrotron Self Compton,
process, SSC see \citealp{jones, marscher-1, ghisellini-2},
{\mta but also by more {\mttt distributed} sources}
outside the jet that produce external inverse Compton (EC)
scattering of seed photons off the highly relativistic electrons
in the jet (e.g., \citealp{sikora-1, paggi}). Such
sources are constituted by the accretion disk around the BH, and by
ionized clouds of gas generally organized in a Broad Line Region
(BLR) {\vv that scatter/reprocess the ionizing radiation from the
disk (see \citealp{ghisellini-3}), and also by the embracing dusty
torus \citep{sikora-2}}. However, {\vv some} interesting new
observations challenge this basic picture, and stimulate
discussions on the nature of the seed photons and on the location
of the \gray {\vv production} site in {\vv FSRQs}.

A first challenging instance {\mta has been} provided by 
 PKS 1830-211 ($z=2.507$) which during a {\mta one-month}
enhancement up to $ 2\times10^{-6}$ \phcmsec {\vv in $\gamma$-rays around 1 GeV}
exhibited a strong flare around 14 October 2010: the
\gray flux reached $ 14\times10^{-6}$\phcmsec in five hours
(\citealp{ciprini} Atel 2943), with very little or no radio, optical
and X-ray counterparts \citep{donnarumma}. This
curious behavior can be understood {\v by} considering
rapid and strong {\mta variations} in the density of external seed
photons as seen by a {\mta relativistically} moving blob or plasmoid {\mta in
the jet} as it approaches the BLR system of clouds.

An  even richer set of  data was provided by the
exceptional flare of the 
 3C 454.3 ($z=0.859$) at GeV energies. {\vv This object entered
a high state on 2 November 2010 maintaining \gray fluxes around $
10\times10^{-6}$\phcmsec for a week. Thereafter the source was
observed by AGILE  and by Fermi-LAT in the energy range 100 MeV -
30 GeV while jumping up by a factor 4 on 17 November 2010, and
reaching the peak flux  $F \simeq 80 \times 10^{-6}$\phcmsec on 20
November 2010}. During this {\vv period}, the source was
extensively monitored in the radio, IR, optical, and X-ray bands (see
\citealp{vercellone-2, vercellone-1, striani, abdo, wehrle, jorstad-2}).
{\vv Models of the high energy emission have been presented} in \citealp{khangulyan} and \citealp{wehrle}
(see also \citealp{marscher-2}).

{\mttt In FSRQs,} $\gamma$-ray emission above 100 MeV often
appears to be correlated with radio emission as shown by the
analysis of \citealp{wehrle}. {\mta On the other hand,
simultaneity of } optical and UV emissions with $\gamma$ rays is often
dubious. {\mttt In particular}, in 3C 454.3 a strong precursor was
observed in the R, optical and UV {\mta bands} on November 10,
2010, i.e., seven days before the \gray jump and ten days before
the \gray peak (see Fig. 2). Unfortunately, radio and IR data were
not obtained during this episode, while the optical has been observed to double its
flux and then fall back in one day. Around 10 November 2010, the
high energy component of the \cc spectrum showed at most a modest
enhancement by $\sim 20$\% in
X-rays, and less or none in $\gamma$ rays.

Remarkably, {\mta  on November 17, 2010} the \gray flux jumped up
by a factor of 4 on a timescale significantly {\mta shorter } than
1 day, whereas the optical and X-ray fluxes increased only by a
factor of 2. 
Moreover, the \gray flux
showed variations by $50\%$ on a timescale of 3 hours, whereas
similar optical and X-ray variations occurred on a timescale of 6
hours (see Figs. 2 and 3). These facts challenge models based on
synchrotron and inverse Compton emission from a homogeneous
region. \textit{Strong and fast} \gray variations with milder or
no counterparts in
X-ray and optical require correspondingly
\textit{strong and fast} {\mta variations} in seed photon {\mta
density as} seen by an emitting region moving along the jet with
 bulk Lorentz factors $\Gamma \sim 10$, as envisioned in
\citealp{vercellone-1}. It is interesting to note that {\mta in
\cc} a remarkable radio plasmoid, {\mta named}  "K10", was
detected {\mta by \citealp{jorstad-2}} to emerge from the radio
core about 160 days after the super-flare. By tracing back the
trajectory of K10 {\mta the authors deduced a {\vv close
starting time of the plasmoid and of}
the $\gamma$-ray flaring activity. This {\mta
finding} supports the view advocated in the present paper that
relativistically moving plasmoids in the jet are {\vv also related to} the
high-energy emission of \ccp.

Furthermore, \citealp{leon-tavares} reported variations of the
MgII emission line in \cc during 
{\mttt the month of } November 2010 (see also \citealp{isler}).
These {\mta observations} suggest that "mirror" {\vvv processes
involving a plasmoid that approaches a specific gas cloud are
relevant to
%
the \gray flare event}; {\vv mirror models
considerably contributing to the local density of seed photons
have been} applied {\mta to clouds } in the BLR by \citealp{ghisellini-3, bottcher-1}.
{\vv In the present paper
we take up the view that mirror effects operate to cause} complex,
multi-frequency behavior {\mtt of \cc} also  at parsec scales,
where \citealp{wehrle} (see also \citealp{jorstad-1}) locate the
high energy emissions of \source.

{\vv In fact,} we will present here a physical model {\vv covering the}
multi-band behavior of \cc  during its exceptional
activity of November 2010, and study the {\mta resulting}
constraints on the \gray emission  processes and location.
In Sect. 2 we discuss some mechanisms that can lead to strong flares
in \gray band by the EC process. In Sect. 3 we model the activity and the flare of
\source in Nov. 2010. Sect. 4 is devoted to discuss our results, and Sect. 5
summarizes our conclusions. The Appendix contains expressions and references that we used in
computing our model.

{\v In the observed physical quantities (timescales, and
radiative fluxes) we include the cosmological redshift effects in
standard form; we adopt the "concordance cosmology": flat geometry with round parameters
$H_0=70\,$km s$^{-1}$Mpc$^{-1}$, and $\Omega _m=0.3$.}


\section{Two scenarios for $\gamma$-ray flaring by EC of cloud photons}

{\vv We consider a multi-zone, time-dependent, leptonic jet model focused on
a schematic and yet effective geometry that bases on relativistic
plasmoids traveling outwards of the compact source of
\cc, and moving toward, through and beyond the BLR shell.}
For the interactions {\vv of plasmoids  with} clouds
we {\vv consider} two {\mtt main}  
scenarios {\vv which} can lead to \gray flaring behavior (see
{\mttt Fig. 1}). Our radiation model is focused  on EC scattering
{\vv to GeV $\gamma$ rays of soft seed photons}; the most relevant
seeds are constituted by the optical-UV photons coming from clouds
in the {\vv main} BLR or beyond, that interact face-on with the
jet electrons.
  {\vv We consider in our calculations seed photons contributed by
the dusty torus and by the accretion disk, entering
from behind and sideways; these seed
photons contribute mostly  to the weaker hard X-ray
radiations \citep{sikora-2, ghisellini-5}.
In our calculations of relevant quantities we
will consider two frames: the laboratory frame (marked by unprimed
symbols), and the co-moving frame (marked by primed symbols).

\begin{figure}
\begin{center}
\includegraphics[width=15cm]{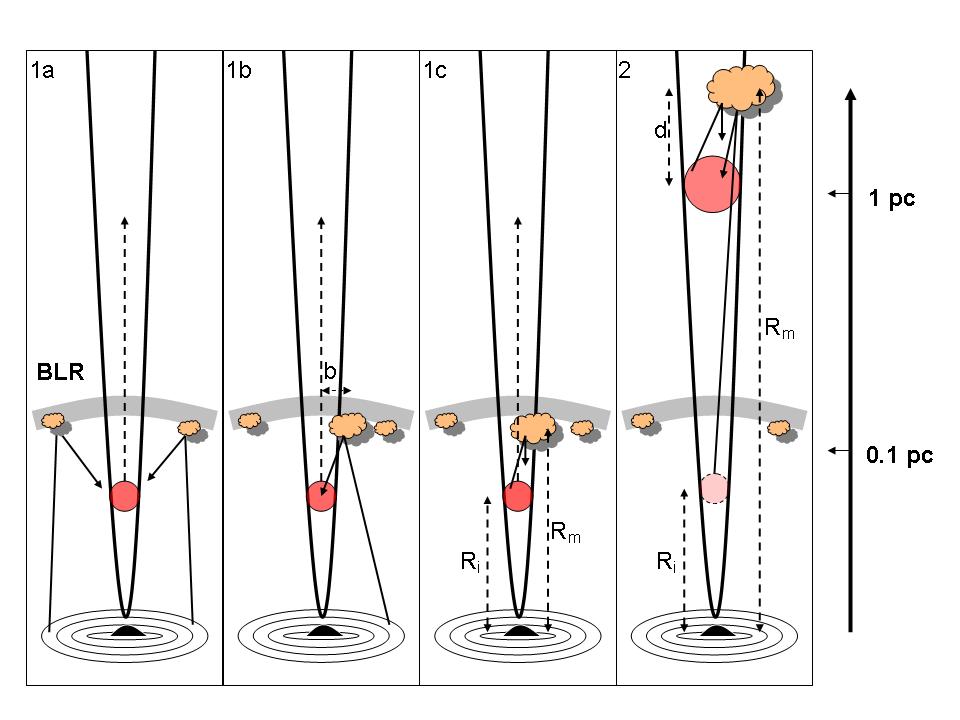}
\caption{Four cases for EC $\gamma$-ray radiation by an outflowing
relativistic {\mta plasmoid}. {\mttt Case 1a}: the {\mta plasmoid}
{\mtt is exposed to the} average
photon field 
{\mtt produced by } the BLR. {\mttt Case 1b:} the {\vv plasmoid is
also} exposed to a locally enhanced photon density as it
approaches a single dense cloud within the BLR {\vv with an impact
parameter $b$}. {\vv Case 1c:} {\mtt the {\mta plasmoid} is
additionally exposed to "mirror" photons produced by itself and
reflected {\mta back} by a single cloud within the BLR}. {\vv Case
2:} {\mtt the {\mta plasmoid} is exposed mainly to mirror photons
reflected by a cloud located beyond the BLR. {\mttt With $d$ we
denote the distance between the mirror cloud and the plasmoid when
the reflected photons re-enter the moving plasmoid
(see Eq. 5).}}
} \label{fig-1}
\end{center}
\end{figure}

\subsection{Case 1: seed photons from clouds in the BLR}

Clouds in a BLR shell surround {\mtt a blazar}  central engine at
a distance $R_{BLR}\sim\,5\times 10^{17}$cm, and cover a small
fraction $a\sim 10\%$ of the solid angle; at larger distances, the
disk becomes inefficient to adequately ionize the clouds {\mttt
(see \citealp{netzer, peterson}).
%
Clouds typically {\mttt scatter/reprocess} the ionizing disk
luminosity $L_D$ and re-emit a fraction  $f\lesssim 20\%$ thereof;
thus an average photon energy
density
\begin{equation}
U_{BLR}\approx a\,f\frac{L_{D}}{4\pi cR_{BLR}^2}
\end{equation}
{\vv is present within $R \simeq R_{BLR}$ (see Fig. 1, Case 1a).}
This provides adequate seed
photons to be upscattered into $\gamma$ rays by
the electrons in a plasmoid of {\vv comoving size scale}
$r_{b}$, itself moving with {\mttt velocity $ v = \beta \, c$ }
and a Doppler factor $\delta = \Gamma^{-1} \, ( 1 - \beta
\cos\theta)^{-1}$ ($\theta$ is the angle of the jet axis
relative to our line of sight,  and $\Gamma = (1 - \beta^2)^{-1/2}
\sim$10 is the bulk Lorentz factor of the plasmoid, see \citealp{begelman}).
In fact, the plasmoid experiences in its
reference frame a boosted field {\mttt $U_{BLR}^{'}\approx (1 +
\beta^2/3)\Gamma^2U_{BLR}$ \citep{dermer-3}.

Correspondingly, fast \gray variability {\mtt can be  observed} on
a timescale $\tau \sim (1\, \rm {\rm day})
(\tau'/10\,days)(\delta/10)^{-1}$, {\mtt where $\tau'\sim r_{b}/c$
is the light crossing time of the plasmoid in the comoving frame.
%
Variability in $\gamma$ rays (and in X rays) is expected to be
produced by EC scattering, and to be closely correlated with the
variability of the synchrotron emission from the jet; as a result,
optical and \gray emissions vary by similar amounts on similar time scales. The
relativistic motion of the plasmoid along the jet causes the \gray
radiation by EC from the BLR seeds to be observed for a time
\begin{equation}
\Delta t_{obs}=\frac{R_{BLR}}{c}(1-\beta\, cos\theta),
\end{equation}
that for viewing angles $\theta\sim \Gamma^{-1}$ {\mtt becomes} 
$\Delta t_{obs} \simeq R_{BLR}/\Gamma^{2}c\sim $ 1 day.
%

Similar expressions also apply  in the case of seed photons 
{\mttt steadily} originating from a dusty torus, on replacing
$R_{BLR}$ with the {\vv torus effective distance} $R_T\sim 1$ pc, and
$L_D\sim10^{46}$erg s$^{-1}$ with the infrared torus luminosity
$L_T\sim10^{-1}L_D$ (e.g., \citealp{ghisellini-5}). {\vv
Detailed treatments of seed photon field anisotropies have been
developed at increasingly sophisticated levels by several authors
({\vv e.g.,} \citealp{bottcher-2, dermer-4, marscher-2, joshi-2}).
{\vv For \gray production the most
relevant seeds are constituted by the optical-UV photons coming
from clouds in the BLR, that interact face-on with the jet
electrons;
here we focus on the strong anisotropies in the comoving frame
introduced by the relativistic motion of the plasmoid
(see Appendix, and \citealp{dermer-2}). We denote this
disposition with "Case 1a" in Fig. 1.


The expression in Eq. 1 was derived {\mttt on assuming that the
BLR constitutes}  a continuous surface that partially reprocesses
the ionizing radiation from a much smaller disk; it describes an
average photon field {\vv which is} nearly uniform within
$R_{BLR}$ and steady on {\mtt a 1-day time scale}. But when a
plasmoid of {\vv comoving radius} $r_{b}\sim 10^{16}$cm approaches
{\mttt with impact parameter $b\sim r_b$} a cloud {\mttt of size
$r\sim 10^{16}$cm}, the \textit{discrete} nature of the BLR {\mttt
emerges}. Strong anisotropy effects (e.g., \citealp{dermer-4})
appear when the plasmoid approaches a specific cloud, either in
the {\vvv canonical} BLR shell (of width $\Delta R_{BLR}$) or
beyond; {\vv therefore, we focus  on the "head-on" geometry, a
condition for which it is reasonable to consider an isotropic seed
photon field.}

{\mttt We take $r_b \sim r\,<<\,\Delta R_{BLR}$} and focus on the
main kinematic and radiative features.} A locally enhanced
radiation field
\begin{equation}
U_{loc}\approx U_{BLR}[1+a^{-1}(\frac{r}{2b})^{2}]
\end{equation}
obtains during the crossing time $\beta^{-1}(r+r_{b})/c $. The
moving blob experiences a boosted field $U_{loc}^{'}\approx
\Gamma^2U_{loc}$. {\vv During the plasmoid-cloud interaction} the
EC radiation leads to observed flux variations by a factor
\begin{equation}
g\equiv 1+a^{-1}(\frac{r}{2b})^{2}\lesssim5
\end{equation}
in the \gray band only; 
{\mta this implies a short } observed timescale
$(r+r_{b})/\Gamma^{2}c \sim\,$ a few hours
(see Eq. 2). The factor $a^{-1}$ in Eqs. 3 and 4 arises from the
fact that now the reflecting cloud obviously covers $\sim100\%$ of
the projected cloud surface $\pi r^2$.

Such a disposition (that we denote Case 1b in Fig. 1) can
explain the enhanced EC radiation in the $\gamma$-ray band during
flares with no correlated counterparts in other bands. In fact,
these flares are likely due to a local enhancement of optical-UV
external photons close to the plasmoid; the observer {\mtt
detects} these photons as scattered by the relativistic electrons
in the plasmoid to energies $\epsilon_{\gamma} \simeq ({\rm 100 \,
MeV}) \, (\Gamma/10) \, (\epsilon_{BLR}/{\rm \ \, eV}) \,
(\delta/10) \, (\gamma/10^3)^2$, where $\gamma$ and
$\epsilon_{BLR}$ are the electron and seed photon energies. On the
other hand, the {\mtt possibility} that a plasmoid meets a cloud
system as described above is confined to distances $R\lesssim
R_{BLR}\sim 5 \cdot 10^{17}$cm from the central black hole, where
dense, disk-illuminated clouds are realistic. We note that the
rate of such plasmoid-cloud encounters is expected to be low just
because of the small BLR {\vv overall} covering factor
$a\sim10\%$.

{\vv The process can produce Compton dominance in events such as
the one from \cc in November 2010  (see \citealp{vercellone-1})},
and in the "orphan" \gray flare of {\mttt PKS 1830-211} in October
2010 \citep{donnarumma}. However, by itself it offers no
explanation for the other, long delayed emissions from \ccp; a
similar conclusion has been reached by \citealp{joshi-2} after a
detailed treatment of the spectral evolution during the plasmoid
outflow toward the BLR.

\subsection{Case 2: seed photons from a mirroring cloud beyond the BLR}

{\vv In this Section we proceed to consider the interaction of a plasmoid with single clouds illuminated
by it.} {\mttt The case of a mirror cloud positioned inside the BLR shell
has been treated by \citealp{ghisellini-3}, and by
\citealp{bottcher-1}, see Fig. 1, {\vv Case 1c}. In this
configuration {\vv a cloud is illuminated by the emission from the
plasmoid itself, and the effective seed field is considerably
contributed by the photons scattered back by the cloud}. The
observed {\vv time lags} between optical emission (originated
closer to the BH)  and the mirror emission are expected to be of
order 1 day only (see Eq. 2). Thus  we  consider a more extended
configuration based on a mirror cloud located \textit{outside} the
BLR ({\vv Fig. 1, Case 2}). This configuration is marked by long
{\vv time lags} and rare occurrence, both fitting the 3C 454.3 features
as we discuss below.}

{\vv Let us then} consider a cloud located at distance $R_m>>R_{BLR}$ from the BH
and {\mttt crossing} the jet outflow. At such distances from
the BH, the disk becomes inefficient {\vv for ionizing clouds
(e.g., \citealp{netzer, peterson})};
on the other hand, {\mttt reprocessing} material with
velocity dispersions $v\approx\sqrt{GM(<R_m)/R_m}\sim
10^3\,-\,10^4$ km$\,s^{-1}$ can still be present there. In such a
case, the illuminating continuum for a moving plasmoid can be
provided in the forward direction by the plasmoid itself; in fact,
its beamed synchrotron emission illuminates {\mttt a} cone with
aperture $\sim \Gamma^{-1}$ around the jet axis.

Specifically, we consider a mirroring cloud placed at a distance
$R_m\sim$ {\vv a few} parsecs along the jet axis, and a plasmoid
with bulk Lorentz factor $\Gamma\sim10$ emitting an intrinsic
{\mta synchrotron} power $L'_{S}$. Owing to the {\mta plasmoid}
relativistic motion, these seed photons (emitted at distance $R_i
< R_m$ from the BH) re-enter {\mta the plasmoid} {\mttt when the
latter } is very close to the mirroring cloud, namely, at a
distance
\begin{equation}
d=\frac{R_m-R_i}{(1+\beta)^2\Gamma^2}\simeq ({\rm 7.5 \times
10^{15} \, cm}) \left( \frac{R_m}{\rm 1 \, pc} \right) \, \left
(\frac{\Gamma}{10} \right)^{-2}, \label{eq-5}
\end{equation}
{\mttt \noindent where we have  taken  into account the causality
constraint pointed out by \citealp{bottcher-1}, that in our
case reads $ R_m - R_i - d = \beta(R_m - R_i + d)$ }.
Thus the plasmoid experiences in its frame a
\textit{doubly} boosted, local photon density, starting from
{\mttt $f \, L'_{S} \, \Gamma^{4}(R_m-R_i)^{-2}/c$ } and suddenly
{\mttt growing up to the limiting value}
\begin{equation}
U_{loc}^{'}\approx \frac{fL'_{S}\Gamma^4}{cd^2}
\end{equation}
when $L'_{S}$ remains closely constant {\mta just {\mttt on}
reaching the mirror} (see {\mtt also} \citealp{ghisellini-3} and
\citealp{bottcher-1} for applications of the process within
the BLR). In the above, we assumed the intervening cloud to fill
the cone illuminated by the beamed {\mta synchrotron} radiation
when the mirror cloud has a size comparable with the jet radius.
{\mttt Such a "head-on" geometry  maximizes the  radiative output
as shown by \citealp{dermer-1}}.

Thus a "mirror-flash"
{\mtt can be} radiated in the \gray band due to EC of {\mttt
reprocessed} photons off the relativistic electrons in the
plasmoid. Eventually, the plasmoid impacts the mirror and radiates
a broad-band synchrotron - EC flare (see Fig. 4 and 5); this will be
observed to closely follow the mirror flash with a time lag
\begin{equation}
\frac{d}{c\Gamma^2}\simeq \frac{R_m}{4 \, c \, \Gamma^4}
\end{equation}
which can be as short as  1/2 hour. On the other hand, the typical
duration of the mirror flash is $\tau \sim
\Gamma^{-2}(d+r+r_{b})/c\sim $ a few hours, {\mttt where $r\sim
10^{16}$cm is the size of the outer mirroring cloud (see Fig. 1,
case 2). Since
we assume 
 $r \sim d$, } the {\mttt mirror} flash
and the {\mttt impact} flare are {\mtt expected} to appear nearly
simultaneous.

A feature supporting this {\vv view} is provided by a
variation of the {\mtt emission line } flux simultaneous with the
optical continuum's variation; an enhancement of $30\%$ in the
lines is expected when the optical continuum from the plasmoid
doubles its flux and ionizes a mirror on its trajectory
(\citealp{ghisellini-3}, see also \citealp{leon-tavares,isler}).
Fig. 6 outlines the spectral evolution associated {\mtt
with the geometry of Case 2.}

\section{Modelling the exceptional $\gamma$-ray flare of 3C 454.3 in November 2010}

The \cc lightcurves observed in different bands (see Figs. 2 and
3) show the activity of \cc to start around MJD 55502, with fluxes
in the optical, X-ray and \gray bands that double in one day
relative to the baseline values. Then the fluxes stay at these
enhanced levels for two weeks (the plateau level of $\gamma$-ray
emission). {\mtt On MJD 55510}, the optical and UV fluxes showed a
remarkable flare with no $\gamma$-ray variation; {\mttt they} then
doubled, and eventually {\mtt fell} back to the previous value in
one day, with no comparable {\mtt variations} in other bands,
except for a moderate enhancement of $20\%$ in soft X rays .

Around MJD 55517, the \gray flux 
{\mtt dramatically jumped up} by a factor {\mtt of $\sim 4$}
while the optical and
X rays increased simultaneously by only a factor {\vv of $\sim 2$}.
{\mttt At later times},  {\mta $\gamma$-ray}, X-ray and optical
fluxes increased together by the same factor, up to a maximum
{\mttt around MJD 55520}, {\vv after which} they  decreased
together. Remarkably, during the entire period} of activity a
significant enhancement of the emission line flux was detected
especially since MJD 55510 as reported by \citealp{leon-tavares} and \citealp{isler};
as we discuss below, this line emission plays an important role in our modelling.
We stress that such a complex behavior is hard to reconcile with standard
co-spatial models of broad-band emissions for the whole stretch of
activity.

Radio maps \citep{wehrle} show that the \gray flare activity
is in apparent coincidence with {\vv the super-luminal knot K10
which is observed to cross the core with size $R_c \approx 16 $
pc}. The plasmoid K10 is resolved on its detaching from the core
$\sim0.5$ yr after the \gray flare. We note that the light
travel-time $R_c/c\sim50$ yr at a viewing angle
$\theta=\Gamma^{-1}$ corresponds to an observed (Doppler
contracted) lag of 0.5 yr for $\Gamma=10$ (see Eq. 2), with an
(apparent) projected super-luminal speed $\sim 10\,c$. {\vv In the detailed
timing of the radio knot allowance should be made for an initial
self-absorbed stage at mm wavelengths\footnote{Detailed fits to
the radio spectra are beyond the scope of the present paper.}.}

\begin{figure}
\begin{center}
\includegraphics[width=17cm]{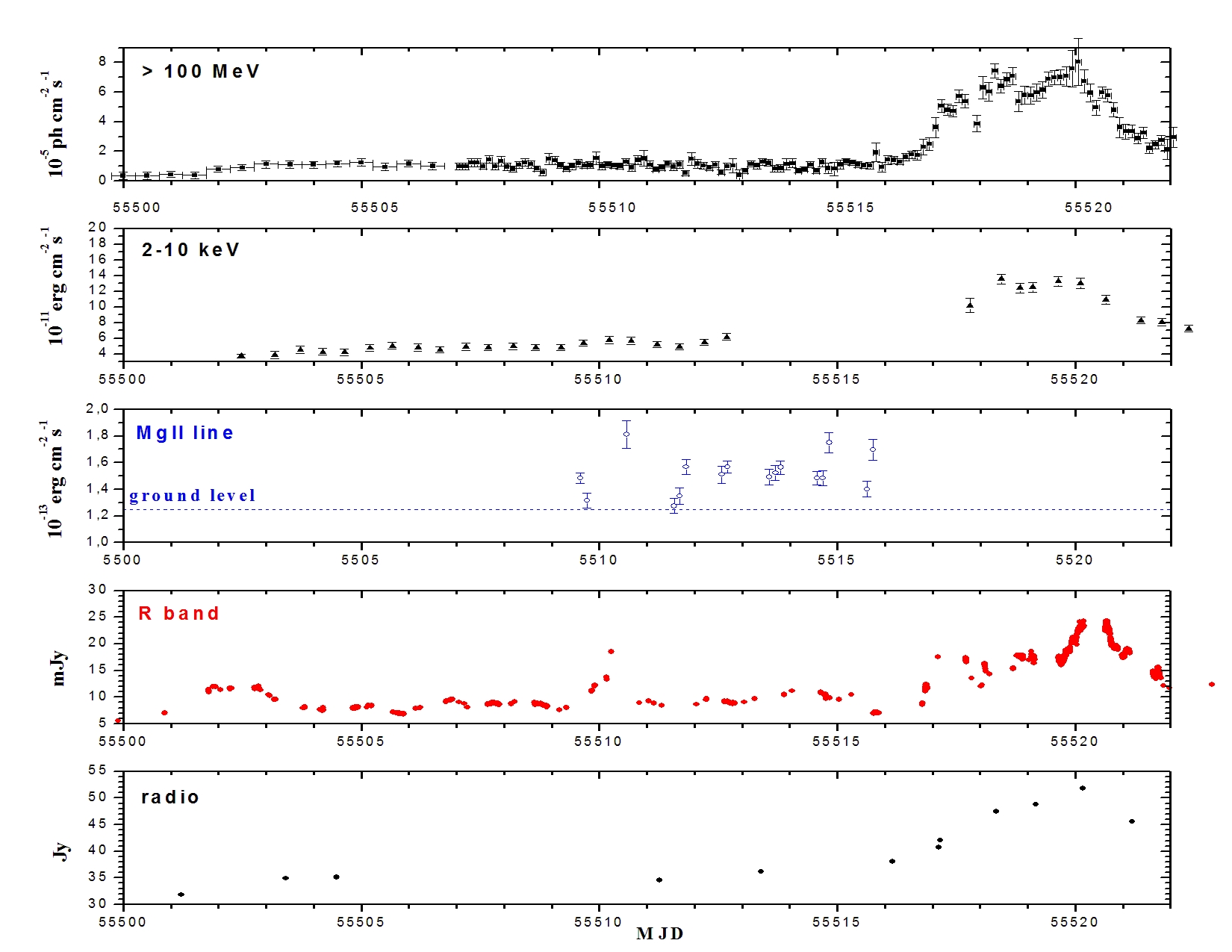}
\caption{Multifrequency monitoring of the blazar 3C 454.3 in
November 2010 ({\mttt MJD 55510 corresponds to November 10,
2010}). The panels show: (1) the $\gamma$-ray emission above 100
MeV as monitored by Fermi-LAT with a time bin of 3 hours. All the
fluxes and spectra were obtained using the Fermi Science Tools,
performing an unbinned likelihood analysis of the publicly
available Fermi-LAT data set. The data analysis includes the
Galactic and isotropic diffuse emission (using
gal-2yearp7v6-v0.fits, iso-p7v6source), and all sources in the
second LAT source catalog within $10^o$ from 3C454.3 (top panel);
(2) X-ray emission in the range 2-10 keV as monitored by Swift and
reported in \citealp{vercellone-1, wehrle} (second
panel); (3) Mg II emission lines  detected by \citealp{leon-tavares}
(third panel); (4) optical GASP WEBT data of R-band
(\citealp{vercellone-1}; fourth panel); (5) radio emission (230
GHz) monitoring reported in \citealp{vercellone-1, wehrle}
(bottom panel). Note the 1-day optical flare around MJD 55510
with no detected counterpart in the X-ray and \gray bands.}
\label{fig-2}
\end{center}
\end{figure}

\begin{figure}
\begin{center}
\includegraphics[width=17cm]{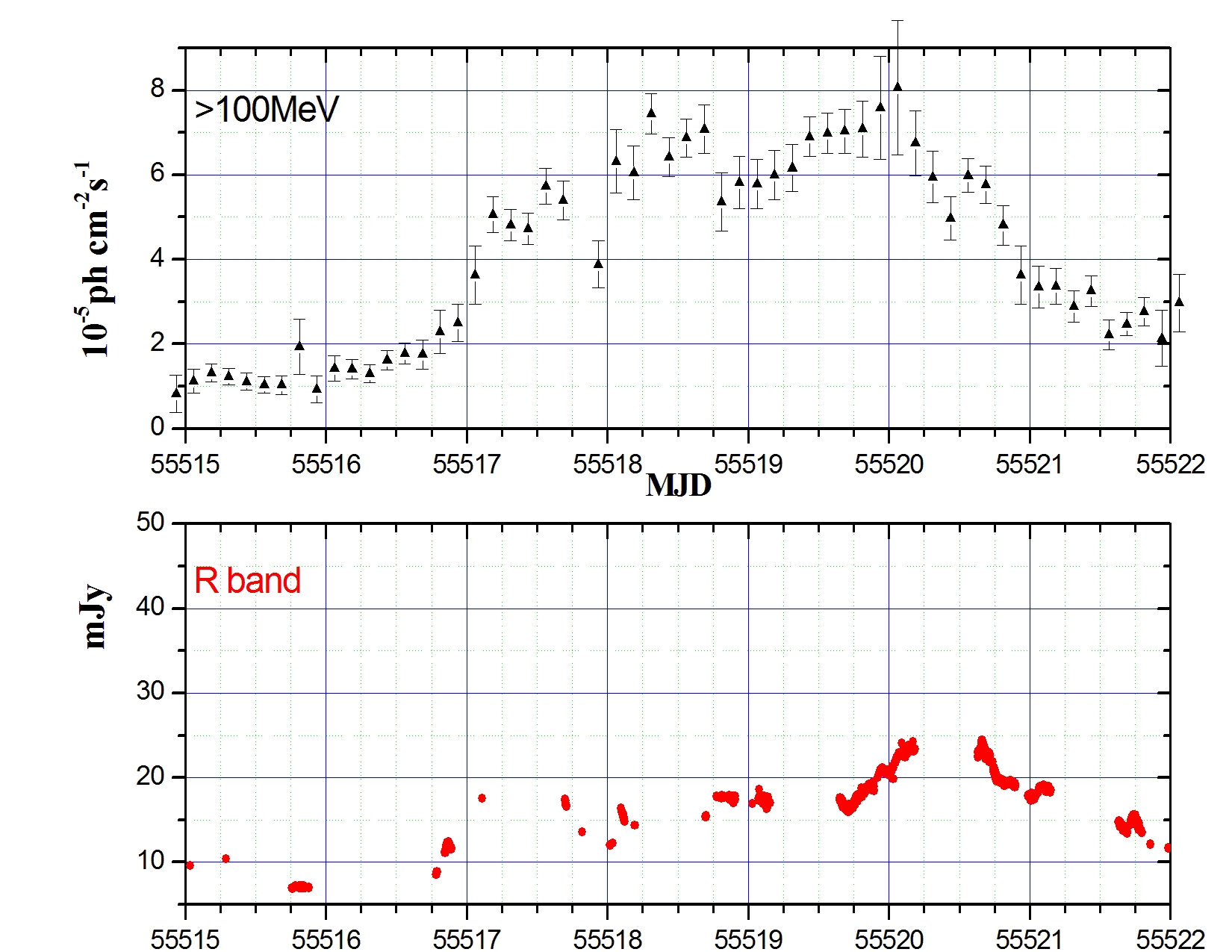} \caption{Multifrequency
monitoring of the blazar 3C 454.3 during the exceptional \gray
flaring in November 2010. Zoom of Fig. 2 showing the
\gray emission above 100 MeV (top panel), and the R-band
(bottom panel) during the peak $\gamma$-ray emission. Note around
MJD 55517 the rise by a factor 4 in $\gamma$ rays, with the
optical rising by a factor of only 2, whereas around MJD 55520 the
flux in both bands rises by the same factor.} \label{fig-3}
 \end{center}
 \end{figure}

\begin{figure}[h!]
\begin{center}
\includegraphics[width=15cm]{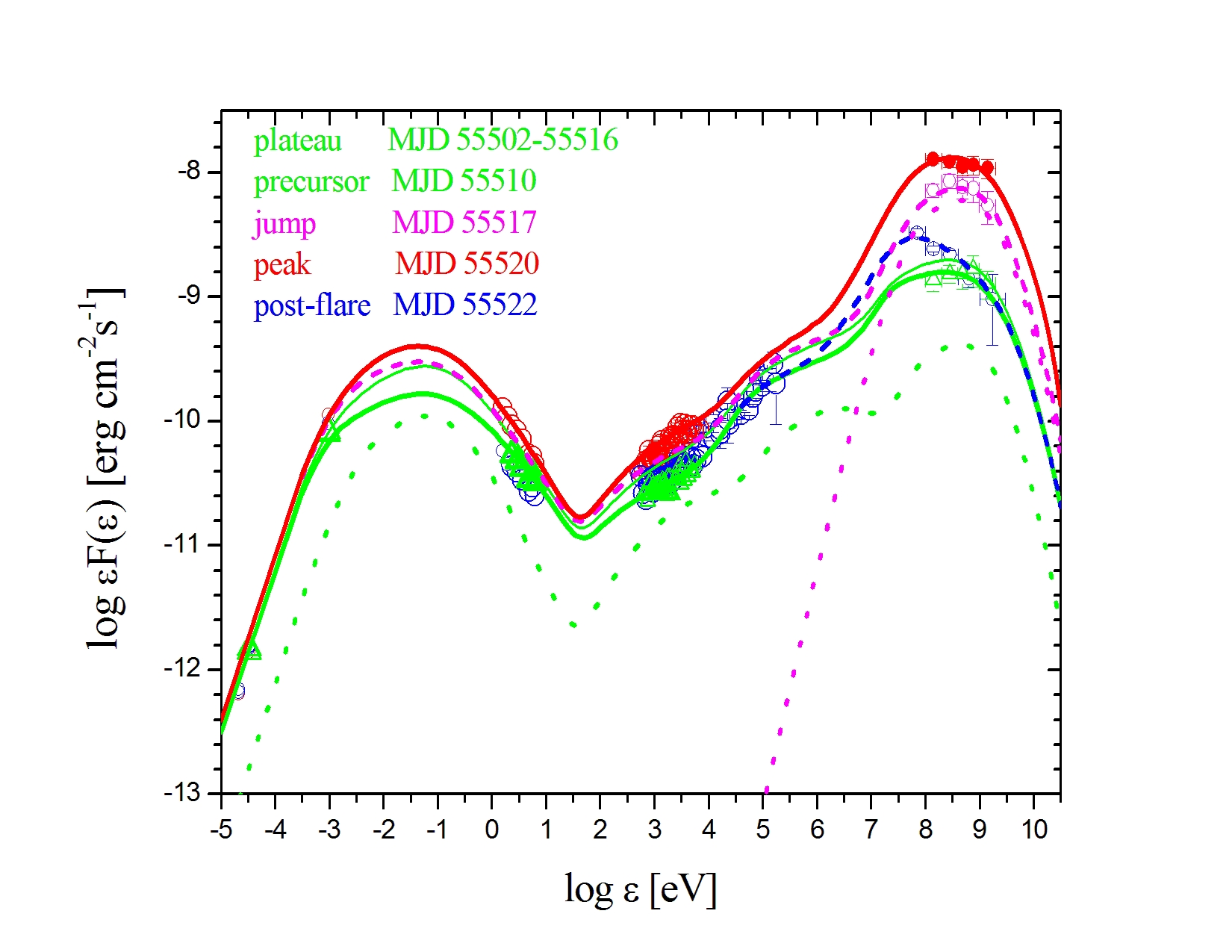}
\caption{Broad-band spectral energy distributions ({\vv SEDs}) computed for
different states of 3C 454.3 {\mttt in November 2010} as indicated in the Appendix
(Eqs. A3, A5, A7 and A9).
The thick green
curve gives the overall plasmoid radiation in the plateau state
{\vv (MJD 55502-55516, Nov. 2-16, 2010). The green dotted curve
shows the radiation of the dominant plasmoid with enhanced optical
emission (MJD 55510, Nov. 10, 2010)}. This component adds
to the plateau state to yield the optical flare SED {\vv (thin, green, solid
line) that   has no} \gray counterpart. The dotted magenta
curve shows the mirror \gray IC radiation. This component summed
with the plateau emission yields the total SED {\mttt on MJD=55517
(Nov. 17, 2010}, dashed magenta line). The thick red curve shows
the peak flaring emission SED {\mttt (MJD 55520, Nov. 20, 2010)}.
The blue dashed curve shows the post-flare SED after two days {\vv of
mirror radiation decrease due to} cooling. Data in the \gray
band are from {\mttt \textit{Fermi}-LAT; all other data are from }
Vercellone et al. (2011). } \label{fig-spectrum}
 \end{center}
 \end{figure}

\begin{figure}[h!]
\begin{center}
\includegraphics[width=15cm]{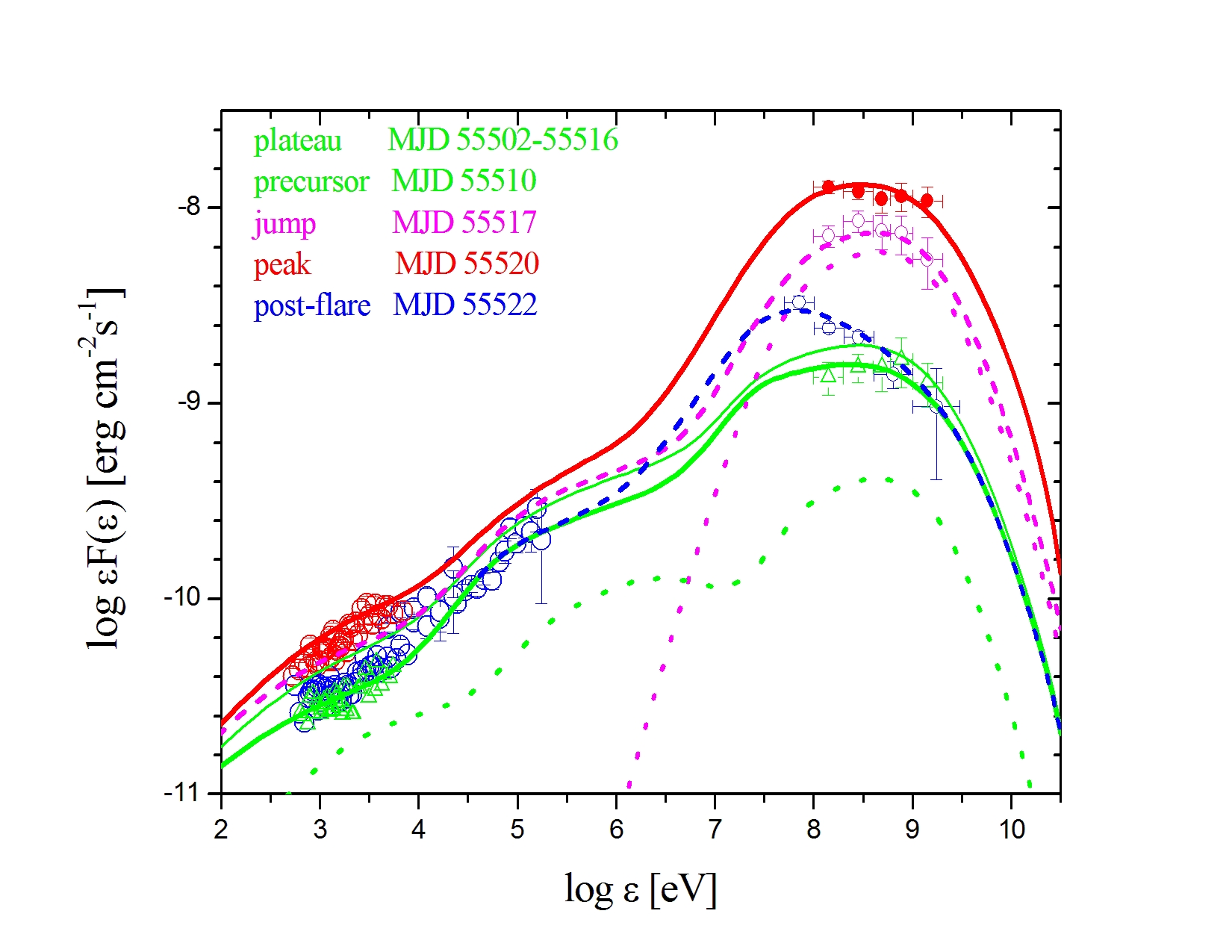}
\caption{Zoom of the Fig. 4 showing the high energy {\vv SEDs} of 3C 454.3
for different states. } \label{figz-spectrum}
 \end{center}
 \end{figure}

Toward understanding the complex multifrequency  behavior of \cc
during the entire activity, we start from the standard picture of
a single plasmoid propagating inside the BLR. On crossing the BLR
the plasmoid produces \gray emission of expected duration $\sim
R_{BLR}/\Gamma^{2}c\sim$1 day (see Eq. 2). At later times, the
plasmoid interacts with the radiation emitted/reflected by clouds
located either close to, or beyond the {\vvv main} BLR as outlined
in Figs. 1 and 6. On the other hand, if we assume production of
relativistic plasmoids continued for at least 2 weeks in the
observer frame, an interesting picture emerges. Such a "plasmoid
train"\footnote{A plasmoid train, i.e., a sequence of density
enhancements, can be provided either by intrinsically intermittent
source activity, or more likely by a sequence of relativistic
internal shocks (as proposed by \citealp{rees}, and developed by many
others including \citealp{joshi-1}). The latter process is
appropriate for heavy jets where the energy outflow is dominated
by the proton kinetic energy $\Gamma m_pc^2$.} {\vv comprising a
{\mttt dozen plasmoids} with a  dominant one addresses both issues
of major flaring events and of the underlying, month-long
plateau.}

{\vv For distances within $R_{BLR}$ these plasmoids steadily produce the observed plateau
with standard synchrotron plus EC spectra as illustrated in Fig. 1, Case 1a.} The EC radiations
off seed photons from the disk, {\mttt the dusty torus} and the BLR
produce enhancements in X rays and $\gamma$ rays by comparable
amounts as the synchrotron rises in the IR, optical and UV bands.
This picture agrees with the observations covering the first two
weeks of plateau activity, i.e., from MJD 55502 to 55516, {\vv see green curves
in Fig. 4 and the first two panels in Fig. 6 relative to Case 1a}.

On the other hand, from a mirror cloud located within, or close to the jet cone at an
height $R_m \approx 2\,$pc, {\vv (Case 2)} we have additional
effects on the radiations from the plasmoid train. {\vv For the
latter to be responsible for the plateau lasting two weeks, the
mirror is to be irradiated by the incoming plasmoids}, and to
re-emit for two weeks in the observer frame. The reflection begins
when the leading edge of the emitting train approaches {\vv the
mirror; in detail, when the distance of the edge from the mirror}
is $(R_m-R_i)\,/\,(1+\beta)\Gamma ^2\simeq 2d$ (recall
from Eq. 5 that $d$ is the distance of the edge when the reflected
photons re-enter the plasmoid).

When the
leading and the following plasmoids reach the mirror and {\mttt crash into} 
it, powerful electron re-accelerations take place {\mttt due to strong
shocks driven by the impacts themselves (see \citealp{khangulyan},
and Appendix) and replenish previous losses. The new injections sustain a
substantial Compton jump in the $\gamma$-ray band relative to the
synchrotron emission detected around MJD 55517. Thereafter, the
electrons radiate by {\mta synchrotron} and by EC in the photon
field established by the mirroring cloud {\vv (see Fig. 6 last two
panels relative to Case 2). When the dominant plasmoid crashes into the mirror}
synchrotron and dominant EC
radiations produce the correlated peaks of similar amounts in
the radio, optical, X rays and $\gamma$ rays that are detected
around MJD 55520. For the production of the optical flare at MJD
55510 we focus on the dominant plasmoid with its enhanced density
and magnetic field that travels within the BLR and emits enhanced
synchrotron radiation without appreciable EC counterparts (as in Case
1a).


Thus we can understand in detail all the time and spectral features of
\cc flare emission within our model including a plasmoid
train that outflows {\vv into a} mirroring cloud located far beyond the BLR.
Concerning the lag of $\tau\simeq15$ days between the start of activity
(MJD 55502) and the sudden jump in $\gamma$ rays at MJD 55517, Eq.
2 leads to an effective travel time (seen at viewing angle
$1/\Gamma$) of $\tau\Gamma^2=1500\,(\Gamma/10)^2$ days, that
corresponds to a length $R_m=1.5\,(\Gamma/10)^2$ pc. Therefore, in
agreement with  \citealp{wehrle}, the peak of the \gray flare
episode is mostly radiated at parsec scales from the central BH.
We locate the plateau emission and the optical precursor closer to
the BH, at the BLR distance or even closer {\vv down to a distance $R_i\sim 10^{17}$cm}.
The high magnetic fields and electron densities expected to prevail at such close distances
lead to enhanced synchrotron emission in the optical-UV}.
These conditions produce the remarkable
{\vv optical flare without} \gray counterpart that is observed around
MJD 55510 as shown in Fig. 2 (see also Fig. 4 and its caption). As anticipated above,
this {\vv in fact requires the dominant plasmoid to have a higher magnetic field
and electron density by some  $20\%$},
relative to other components of the {\mttt plasmoid train} (see
Table 1). The EC counterpart in X-ray and \gray
{\mttt is not easily detectable in the presence of the relatively
{\mtt strong} plateau emission} discussed at the beginning of
this Section.


{\vv The Appendix provides details of our spectral computations.
Table 1 summarizes the parameters of
our modeling which covers the plateau, the flare, and the
post-flare phases. Additional parameters are constituted by the disk luminosity
$L_D=5\times 10^{46}$erg$\,$cm$^{-2}$s$^{-1}$,
the injection distance $R_i=5\times 10^{16}$cm, the BLR radius $R_{BLR}\simeq 10^{18}$cm,
the dusty torus infrared luminosity $L_T=10^{-1}L_D$, the  torus effective distance
$R_T=1\,$pc; the mirror distance is taken $R_m=2\,$pc and its size $r=10^{16}$ cm.
Figs. 4 and 5 show the resulting broad-band
spectral energy distribution as a function of photon energy.}

\begin{table*}[h!]
\begin{center}
\caption{Model parameters for the November 2010 \gray flare of \source.}
 \small \noindent
 \vskip 0.2 cm
\begin{tabular}{|c|c|c|c|c|c|c|c|c|c|}

  \hline
  \bf{Component} &  $\bf{\Gamma}$ & \bf{B (G)} &
  $\bf{r_{b} (cm)}$ & $\bf{K\,(cm^{-3})}$ &
  $\bf{\gamma_b}$ & $\bf{\gamma_{min}}$ & $\bf{\zeta_1}$ & $\bf{\zeta_2}$\\
  \hline
  {\mttt Dominant plasmoid}  &  12 & 1.1 & $4\times10^{16}$ & 50 & $6\times10^2$ & 80 & 2.0 & 4.4\\

  \hline
  {\mttt Subdominant plasmoids}     &  15 & 0.9 & $3\times10^{16}$ & 40 & $7\times10^2$ & 45 & 2.4 & 4.2\\

  \hline
\end{tabular}
\end{center}

{\small \noindent The Table provides the main physical parameters
of the emitting plasmoids responsible for the emissions plotted in Figs. 4 and 5.
The columns give  the bulk Lorentz factor $\Gamma$, the average comoving
magnetic field $B$, and the comoving radius $r_b$ {\vv for the dominant and for
the average plasmoid in a train of a dozen.} The subsequent columns
give for the associated electron distribution functions
$n_e(\gamma)$: these are modeled as broken power-law with
the normalization $K$, the minimum particle energy $\gamma_{min}$, the break $\gamma_b$, and
the low- and high-energy indices $\zeta_1$ and $\zeta_2$ (see Appendix).}
\vspace*{0.5cm}
\end{table*}
\normalsize
\bigskip

{\vv Here we stress that our computed spectra agree with the observed overall and specific
features.
In particular, the slope $\zeta_2$ specifically reflects into the steep shape of the
optical - UV continuum, and the flatter slope $\zeta_1$ into the relatively flat shape
in $\gamma$ rays in the range 100 MeV - 1 GeV.

Note that the low-energy  (and initial)  slope  $\zeta_1$
indicated by the spectral fits and reported in Table 1 turns out
to be flatter in the dominant compared to the subdominant
plasmoids, (see also Fig. 7 in the Appendix).  Higher values of
$\zeta_1$ and density are expected (see, e.g., \citealp{longair}) for
\textit{stronger} shocks driven in the dominant relative to the
subdominant plasmoids. The electron accelerations are easily fed
by the large  bulk energy in the jet and the associated plasmoids,
namely,  by $\Gamma \, m_p \, c^2 >> \gamma_b \, m_e \, c^2$. We
add two points: (i) higher densities and magnetic fields hold
behind stronger shock fronts in  trans-relativistic plasmas; (ii)
lower values of $\Gamma$ coupled with higher magnetic fields agree
with the minimum power arguments for blazar jets down to the core
as discussed by \citealp{ghisellini-1}. These features
consistently support our model.}


%

\begin{figure}[h!]
\begin{center}
\includegraphics[width=17cm]{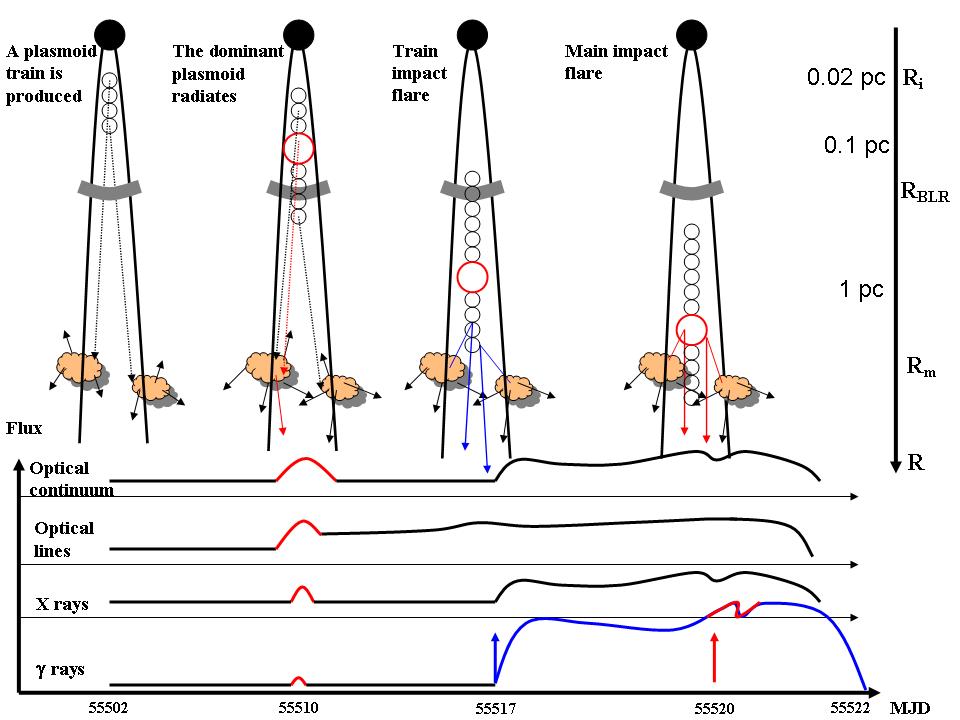}
\caption{Schematic representation of the different steps of
interaction and corresponding emissions in 3C 454.3 after the
crashed-mirror scenario proposed in Sects. 2 and 3. Red color
marks the contributions to the fluxes of the {\vv dominant}
plasmoid to the fluxes, whereas blue color marks the EC-mirror
contributions to $\gamma$ rays of the other plasmoids. The two
arrows mark the times when mirrored photons re-enter the {\vv
leading plasmoid} and the {\vv dominant one, respectively. The
final stage in the figure corresponds to the dominant
plasmoid crashing into the mirror, before the final decay}.} \label{fig-4b}
 \end{center}
 \end{figure}

\newpage
\section{Discussion}

Our model {\mta aims} at a {\mttt specific}  
understanding of {\mttt all} multifrequency emissions of \cc
during its record-high \gray emission episode in November 2010.
The model considers plasmoid production along the jet and their propagation into environments with widely
different densities of soft photons, that scatter off the GeV electrons within the plasmoids.
We find that it is hard to explain all 
{\mta observed features} of the flaring \cc within
{\mta a sparsely covering BLR. Instead,}
 mirroring by a cloud 
 {\mta located along } the jet
trajectory can produce {\mttt locally} a strongly enhanced seed
photon field {\mta yielding an intense EC flare}.

We have followed the evolution of the dominant plasmoid in the train as it
outflows from the inner region up to the distant mirror. Its
{\mttt appearance} is observed in our frame after about a week
from the beginning of the plateau 
due to the plasmoid train. The \gray peak at MJD 55520 is
interpreted as mirror-enhanced EC radiation once the {\mttt
dominant} plasmoid reaches the mirror region, approximately after
10 days in the observer frame.

Fig. 6 summarizes our model and its outputs, spelled
out as follows. To start with, {\mttt a train of several }
plasmoids is {\mttt produced} and {\vv flows  towards and}
throughout the BLR. The resulting synchrotron and EC radiation
{\mttt inside and at} the BLR shell produces the {\mttt
quasi-steady} plateau stretch.
{\vv Plasmoids that leave the BLR are replaced by the trailing
ones  in the train}. The plateau emission therefore lasts some 15
days, with {\mttt wiggles} reflecting internal activity in the
plasmoid train. At some point, a {\mttt \it dominant} plasmoid
occurs; this yields the "orphan" optical flare after about 1 week in
the observer's frame and the enhanced flux of emission lines as
discussed at the end of Sect. 2.2. Eventually, the leading edge of
the train reaches a mirror cloud located well \textit{beyond} the
BLR; strong radiation of the EC type is produced at this point as
explained in detail in Sect. 3. {\vv Evidence of mirror emission
is supported} by enhancement in line-emission and by strong
Compton dominance, {\vv while} retaining a conservative value of
$\Gamma\approx 15$ for the bulk Lorentz factor. While
Compton-dominated radiation continues on, synchrotron emission
adds as ensuing plasmoids crash into {\mttt the outer}
\textit{mirror} cloud. The \gray radiation is still {\mttt
mirror-enhanced}, while increased synchrotron emission is
produced. {\mttt When} the {\mttt dominant} plasmoid reaches the
mirror, a strong peak in both fluxes ensues. Later on, the overall
radiation {\mttt decreases until the
plasmoid production ends, or its {\mttt outflow} is terminated. }


Previous authors \citep{ghisellini-3, bottcher-1}
discussed {\mttt similar mirror} processes within the BLR {\mta at
distances less than $5\times 10^{17} \, \rm cm$}; we find this
location {\vv not adequate} for the events of \source in November 2010. {\mta In fact,} in
interpreting the "orphan" optical flare a clear difference {\mta
arises } between {\mta models based on mirrors placed inside and
\textit{beyond}   the BLR}.
If the {\mttt mirror were located within the BLR,}
the absence of a corresponding increase
of $\gamma$ rays {\mttt would be} problematic. The {\mta enhanced}
lines emitted in {\mta approximate}  coincidence
with {\mttt the}  "orphan" optical flare are 
{\mta indicative} of a 
reprocessing cloud; it is unlikely for this cloud {\mttt to re-emit}
 within the BLR
{\mttt without overproducing $\gamma$ rays.}

{\mta {\mttt On the other hand,} we propose in Sect. 2} that
mirror enhanced EC radiation at
{\mta some}  distance 
{\mta beyond} the BLR {\mta constitutes}  a good candidate to
explain {\mttt \textit{all } known multifrequency and timing
features of \ccp; specifically it explains the time lag of 10 days between the
"orphan" optical  flare and the peak in $\gamma$ rays}. We envisage
the following sequence (see Fig. 6): (1) production of a plasmoid train; (2)
outflow into the BLR {\mttt
leading to}  plateau emission; (3) outflow of the train  
{\mttt beyond} the BLR; (4) interaction with a mirror cloud.
The {\mttt appearance of the dominant} plasmoid  corresponds
to the optical flare without \gray counterpart;
the delayed, strong $\gamma$-ray radiation 
agrees with {\mttt the plasmoid } outflow and interaction {\mttt with
the mirror cloud}.
A virtue of our model is to 
 {\mttt require a
conservative} value of the jet bulk Lorentz factor $\Gamma \approx
15$, a value not stressing the jet acceleration processes
(e.g., \citealp{camenzind}) 
yet consistent with the {\mttt observed} rapid
onset of the $\gamma$-ray flare in November 2010.
{\mta The requirement of}  substantial high-energy radiation from
the jet at large distances from the central BH is indeed the main
feature of our model that {\mttt marks it out } from other
investigations focused on outer emissions but with low radiative cooling {\vv
(e.g., \citealp{ghisellini-4}). 
An added bonus of our proposed disposition is to place most of the
$\gamma$-ray emission at distances $R>3\times 10^{17}cm$, where
the opacity for $\gamma$-$\gamma$ interactions is safely low 
(e.g., \citealp{ghisellini-3}).}

Note that our model requires a sequence of two infrequent events that is
consistent with the rare occurrence (about 1 in several years) of
$\gamma$-ray superflares among the {\mttt whole } population of
blazars monitored by AGILE and Fermi-LAT  {\v (\citealp{pittori},}   http://fermi.gsfc.nasa.gov/ssc/data/access/lat/msl\verb"_"lc).
%

\section{Conclusions}

{\vv We have discussed a set of possible models for }
the exceptionally intense {\mta and complex} 
flare from the blazar \cc in November 2010. We {\mta  have
focussed and developed}
a consistent model {\mta  that}  addresses all existing
observations in different bands from radio waves to GeV {\mta
energies}.
This is based on a train of plasmoids {\mta propagating along
the jet and interacting with the BLR and with the external
environment}, which naturally explains both the 
extended plateau and 
the  intense \gray flare. The {\mta strongly
Compton-dominant} nature of the {\mta latter} 
indicates that EC from a blob crashing into a mirror is at work. 
Difficulties with a mirror cloud 
{\mta  located} within the BLR are overcome by our proposed {\mttt
alternative location beyond the {\vvv canonical} BLR shell}.

{\mttt In fact,} our preferred picture for \cc event is based on a
{\mta location} well 
{\mta beyond} the BLR to produce the \gray super-flare episode.
{\mttt The \textit{sharp} onset (a few hours) of the super-flare
and its strong Compton dominance are specific features of our
model. Mirroring implies a double Lorentz boost that strongly
affects the radiation. On the other hand, the \textit{long}
duration of the $\gamma$-ray enhanced emission is accounted for by
the large distances (of order few pc) implied in our model.}
%
%
So the latter is characterized by {\vv intense "far-yet-fast"
\gray radiation, and can also be applied to similar flaring episodes} other sources {\mta
such as the activity of PKS 1830-211 in October 2010}.

\vskip .1in

{\bf Acknowledgments: } This work is based on satellite data from
AGILE and {\vvv \textit{Fermi}-LAT}, and on ground-based data
taken and assembled by the WEBT collaboration and stored in the
WEBT archive at the Osservatorio Astrofisico di Torino - INAF
(http://www.oato.inaf.it/blazars/webt/). We acknowledge
constructive comments by our referee that helped us to improve our
presentation.  We thank P. Romano for the Swift/XRT data, C.M.
Raiteri, and M. Villata for the GASP/WEBT radio/optical data.{\vvv
We also thank  J. Leon-Tavares for providing optical line data.
Investigation carried out with partial support by the ASI grant
no. I/028/12/0 and I/028/12/2.}

{}

\setcounter{equation}{0}

\renewcommand{\theequation}{A\arabic{equation}}

\newpage
\section*{Appendix}

{\vv We assume the emitting electron population to emerge from an
injection/acceleration stage, provided by shocks (see \citealp{longair}),
with a comoving distribution of the random relativistic
energies $(\gamma m_e c^2)$ in the form of a single power-law;
this is subsequently reshaped into a broken power-law
\begin{equation}
n_e(\gamma)=\frac{K\,\gamma_b^{-1}}{
(\gamma/\gamma_b)^{\zeta_{1}}+(\gamma/\gamma_b)^{\zeta_{2}}}
\end{equation}
with parameters tabulated in Table 1.

These electrons emit a primary
synchrotron spectrum and a second contribution (SSC) produced by
inverse Compton (IC) as the primary synchrotron photons scatter
off the same electron population. We numerically compute the fluxes from these processes
following standard relations given in detail by, e.g., \citealp{blumenthal}, Eqs. 4.52 and 2.61.

For electrons in a magnetic field $B$, the synchrotron SED peaks in the laboratory frame at
\begin{equation}
\epsilon_s=3.7\times10^6\,h\,B\,\gamma_b^2\,\delta ,
\end{equation}
where $h$ is the Planck constant and $\delta=[\Gamma(1-\beta\,
\cos\theta)]^{-1}$ is the Doppler factor due to the bulk flow of
emitters toward the observer at an angle $\theta$ relative to the
line of sight. The SED at the synchrotron peak scales as
\begin{equation}
\epsilon_s\,F(\epsilon_s)\propto \delta^4R^3B^2K\,\gamma_b^2.
\end{equation}
\noindent As to the IC component, its SED contribution peaks at
\begin{equation}
\epsilon_c=\frac{4\gamma_b^2\epsilon_s}{3}
\end{equation}
\noindent with a peak value scaling in the Thomson regime  as
\begin{equation}
\epsilon_c\,F(\epsilon_c)\propto \delta^4R^4B^2K^2\gamma_b^4
\end{equation} if the scattering takes place  with a density of
target photons scaling as $n_{ph}\propto F_s R/c$. The
relativistic motion toward the observer amplifies the observed
power by the factor $\delta^4$, and allows it to vary on a
timescale $ R/c \, \delta$.
%

Fluxes emitted by EC process are computed as given in detail by
e.g., \citealp{reynolds}, (Eq. 12). Here target photons additional to SSC are
provided by sources external to the jet. The high-energy
component of the spectra is contributed by the electrons that
Compton upscatter the external photons. The SED
now peaks at energies
\begin{equation}
\epsilon_c=\frac{4\gamma_b^2\epsilon'_{p}\delta}{3}
\end{equation} and the corresponding peak value scales as
\begin{equation}
\epsilon_c\,F(\epsilon_c)\propto
\delta^4R^3K\,\gamma_b^2N'_{p}\epsilon'_{p} .
\end{equation}
\noindent Here two new ingredients enter: $\epsilon'_{p}$ and
$N'_{p}$ i.e., the peak energy and the density at the peak of the
external photons as seen by the moving plasmoid. These quantities
are related to the energy density $N(\epsilon)$ in the laboratory
frame by means of the bulk Lorentz factor $\Gamma$, in a manner
that depends on the geometry of the system that causes an
additional dependence on $\Gamma$. \citealp{dermer-2}
discuss observed SED dependencies on $\Gamma$ varying from $
\Gamma^3$ to $\Gamma^6$, for photons entering the moving plasmoid
from behind or from the front, respectively. Here we assume for
$N(\epsilon)$ black body distributions at temperatures {\v
$T_D=3\,10^4\,$K, $T_T=3\,10^2\,$K and $T_{BLR}=5\,10^4\,$K }
for the accretion disk, the dusty torus and the BLR, respectively.
{\v Since the peak energy of the external photons in the observer-frame
satisfies $\epsilon_{p}<5\,$eV, the scattering occurs in the Thomson regime; in fact, with the
parameters reported in Table 1 $\gamma\Gamma\epsilon_{p} << mc^2$ results.
If the scattering happened in the Klein-Nishina regime, further steepening of high energy spectra
would be observed.}

By synchrotron and Compton losses the electrons cool down on a comoving
timescale
\begin{equation}
\tau'_{cool}(\gamma)\equiv -\frac{\gamma}{\dot{\gamma}}=
\frac{3m_e c}{4\beta^2\sigma_T\gamma(U'_B+U'_e)},
\end{equation} where $\sigma_T$ is the Thomson cross section,
$U'_B=B^2/8\pi$ and $U'_e$ are the comoving energy densities
before scattering of the magnetic field and of the external
radiation, respectively. This sets a cooling break at the energy
$\gamma_{cool}=3m_e c^2 / 4\sigma_T\, R\,\beta^2(U'_B+U'_e)$,
beyond which the electrons cool down rapidly.

  Focusing on the 
 radiative cooling,  the electron energy distribution for a single plasmoid
(Eq. A1)  evolves according to
\begin{equation}
\frac{\partial \, n_e(\gamma,t)}{\partial t} + \frac{\partial \,
[\dot{\gamma} \, n_e(\gamma,t)]}{\partial \gamma} =  n_{e
\,i}(\gamma_{i})\,\delta(t-t_{i}).
\end{equation} \noindent The results for the dominant (red) and a
subdominant (green) plasmoid are shown in Fig. \ref{fig-7} which
illustrates the detailed effects of cooling on the distributions
after 1 day in the observer frame. The contributions of one
dominant and of several other subdominant plasmoids at the
appropriate stages of evolution are summed  to produce the spectra
of Figs. \ref{fig-spectrum} and \ref{figz-spectrum}. We
schematically illustrate in Fig. \ref{fig-4b} our procedure for
summation and composition of the received fluxes in four spectral
bands.}

\newpage

\begin{figure}[h!]
\begin{center}
\includegraphics[width=17cm]{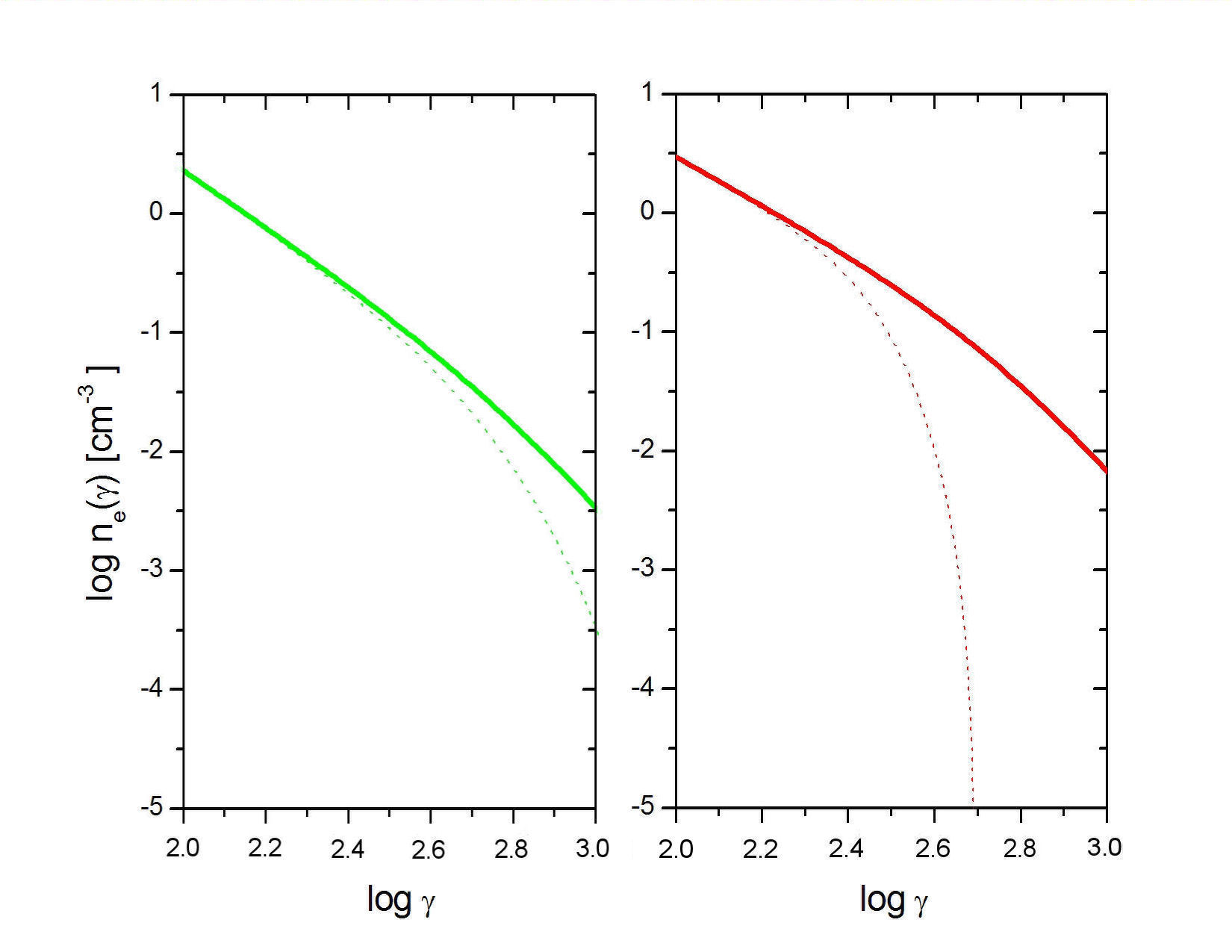}
\caption{{\vv The electron distribution functions evolving  by
radiative cooling in the dominant (red) and a subdominant (green)
plasmoid. The color code corresponds to the spectra in Figs. 4 and
5; thick lines indicate the initial distributions, and thin lines
the evolved ones after 1 day in the observer frame. {\it Left
panel:} we show  the electron population of a subdominant plasmoid
that is injected at $R_i$ within the BLR, and produces the plateau
emission. {\it Right panel:} the electron population in the
dominant plasmoid as it crashes into the outer mirror and produces
the peak on MJD 55520. Note that the initial functions are
somewhat steeper in the BLR as a result of weak internal shocks,
and somewhat flatter at the crash into the external mirror as a
result of stronger shocks.} } \label{fig-7}
 \end{center}
 \end{figure}

\bibliographystyle{apj}


\end{document}